%% file: main.tex
\documentclass[journal]{IEEEtran}

\usepackage{hyperref}
\usepackage{graphicx}
\usepackage{amsmath,amssymb} 
\usepackage{color}
\usepackage{times}
\usepackage{epsfig}
\usepackage{tabularx}
\usepackage{booktabs}
\usepackage{multirow}
\usepackage{array}
\usepackage{xspace}
\usepackage{balance}
\usepackage{enumitem}
\usepackage{mathtools}
\usepackage{rotating}
\usepackage{url}
\usepackage{tabulary}
\usepackage[table,dvipsnames]{xcolor}
\usepackage{subcaption}
\usepackage[export]{adjustbox}
\usepackage{ctable}
\usepackage{diagbox}

\def\Vec#1{{\boldsymbol{#1}}}
\def\Mat#1{{\boldsymbol{#1}}}

\newcommand{\fig}{{Fig.}\@\xspace}
\newcommand{\tab}{{Table}\@\xspace}
\newcommand{\ie}{{i.e.}\@\xspace}

\newcommand{\etal}{{\it et~al.}\@\xspace}

\graphicspath{{./figs/}}

\begin{document}
\title{Video Storytelling: Textual Summaries for Events}

\author{
	Junnan~Li,
	Yongkang~Wong,~\IEEEmembership{Member,~IEEE,}
	Qi~Zhao,~\IEEEmembership{Member,~IEEE,}
	Mohan~S.~Kankanhalli,~\IEEEmembership{Fellow,~IEEE}
	\IEEEcompsocitemizethanks{
		Junnan~Li is with the Graduate School for Integrative Sciences and Engineering, National University of Singapore (email: lijunnan@u.nus.edu).
		Qi~Zhao is with the Department of Computer Science and Engineering,
		University of Minnesota (email: qzhao@cs.umn.edu).		
		Yongkang~Wong and Mohan~S.~Kankanhalli is with the School of Computing, 
		National University of Singapore (email: yongkang.wong@nus.edu.sg, mohan@comp.nus.edu.sg).
	}
	\thanks
	{
		Manuscript received  XXX, 201X; revised XXX, 201X.
	}

}

\markboth{A SUBMISSION TO IEEE TRANSACTIONS ON MULTIMEDIA}%
{Shell \MakeLowercase{\textit{et al.}}: Bare Demo of IEEEtran.cls for IEEE Journals}

\maketitle
\input{sec_abstract}

\IEEEpeerreviewmaketitle

\input{sec_introduction}

\input{sec_literature} 
\input{sec_method} 
\input{sec_dataset}

\input{sec_experiment}

\input{sec_conclusion}

\ifCLASSOPTIONcompsoc
  \section*{Acknowledgments}
\else
  \section*{Acknowledgment}
\fi
\input{sec_acknowledgement}

\ifCLASSOPTIONcaptionsoff
  \newpage
\fi

\balance
\bibliographystyle{IEEEtran}
\bibliography{bib}

\end{document}

%% file: sec_abstract.tex
\begin{abstract}

Bridging vision and natural language is a longstanding goal in computer vision and multimedia research.
While earlier works focus on generating a single-sentence description for visual content,
recent works have studied paragraph generation.
In this work, we introduce the problem of video storytelling,
which aims at generating coherent and succinct stories for long videos.
Video storytelling introduces new challenges, 
mainly due to the diversity of the story and the length and complexity of the video.
We propose novel methods to address the challenges.
First, we propose a context-aware framework for multimodal embedding learning,
where we design a Residual Bidirectional Recurrent Neural Network to leverage contextual information from past and future.
The multimodal embedding is then used to retrieve sentences for video clips.
Second, we propose a Narrator model to select clips that are representative of the underlying storyline.
The Narrator is formulated as a reinforcement learning agent which is trained by directly optimizing the textual metric of the generated story.
We evaluate our method on the Video Story dataset,
a new dataset that we have collected to enable the study.
We compare our method with multiple state-of-the-art baselines and show that our method achieves better performance, 
in terms of quantitative measures and user study.

\end{abstract}

%% file: sec_introduction.tex
\section{Introduction}
\label{sec:introduction}

Generating natural language descriptions for visual content has become a major research problem for computer vision and multimedia research.
Driven by the advent of large datasets pairing images and videos with natural language descriptions,
encouraging progress has been made in both image and video captioning task. 
Building on the earlier works that describe images/videos with a single sentence, 
some recent works focus on visual paragraph generation which aims to provide detailed descriptions for images/videos~\cite{Krause_2017_CVPR,Liang_ICCV_2017,Rohrbach_GCPR_2014,Yu_CVPR_2016,Ranjay_ICCV_2017,Xu_TCSVT_2018,Liu_Access_2018,Liu_CVIU_2017,Park_NIPS_2015,Liu_AAAI_2017}.

Existing literature on visual paragraph generation can be divided into two categories:
a) fine-grained dense description of images or short videos~\cite{Krause_2017_CVPR,Liang_ICCV_2017,Rohrbach_GCPR_2014,Yu_CVPR_2016,Ranjay_ICCV_2017,Xu_TCSVT_2018,Liu_Access_2018,Liu_CVIU_2017} and b) storytelling of photo streams~\cite{Park_NIPS_2015,Liu_AAAI_2017}.
However, other than images and short videos clips, consumers often record long videos for important events, such as birthday party or wedding.
In order to assist users to access long videos, 
previous works on video summarization focused on generating a shorter version of the video~\cite{Dufaus_ICIP_2000,Goldman_ACMTG_2006,Gygli_CVPR_2015,Lu_CVPR_2013,Potapov_ECCV_2015,Zhang_CVPR_2016,Song_CVPR_2015}.
We argue that compared to visual summary,
a textual story generated for a long video is more compact, semantically meaningful,
and easier to search and organize.
In this paper,
we introduce a new problem: visual storytelling of long videos.
Specifically, 
we aim to compose coherent and succinct stories for videos that cover entire events.

The proposed video storytelling problem is different from the aforementioned visual paragraph generation works. 
First, 
compared to dense-captioning~\cite{Krause_2017_CVPR,Liang_ICCV_2017,Rohrbach_GCPR_2014,Yu_CVPR_2016,Ranjay_ICCV_2017}, 
we focus on long videos with more complex event dynamics while not aiming to describe every detail presented in the video.
Instead, 
we focus on extracting the important scenes and compose a story.
Second,
video storytelling is a more visually-grounded problem.
This is in contrast with the storytelling of photo streams~\cite{Park_NIPS_2015,Liu_AAAI_2017},
where the datasets~\cite{Park_NIPS_2015,Huang_NAACL_2016} only consist of a few photos per story.
As a result, 
the annotated stories may depend on annotators' imagination and prior knowledge, 
and the challenge is to fill in the visual gap between photos.
On the other hand, 
the challenge for video storytelling is not lack of visual data,
but to compose a coherent and succinct story from abundant and complex visual data.

The task of video storytelling introduces two major challenges.
First, compared to single-sentence descriptions, long stories contain more diverse sentences, 
where similar visual contents can be described very differently depending on the context. 
However, the widely-used Recurrent Neural Network (RNN) based sentence generation model tends to produce generic, repetitive and high-level descriptions~\cite{Huang_NAACL_2016,Dai_ICCV_2017,Li_HLT_2016}.
Second,
long video usually contains multiple actors, multiple locations and multiple activities.
Hence it is difficult to discover the major storyline that is both coherent and succinct.

\input{table/fig_example}

In this work, we address these two challenges with the following contributions.
First, we propose a context-aware framework for multimodal semantic embedding learning.
The embedding is learned through a two-step \textit{local-to-global} process.
The first step models individual clip-sentence pairs to learn a local embedding.
The second step models the entire video as a sequence of clips.
We design a Residual Bidirectional RNN (ResBRNN) that captures the temporal dynamics of the video and incorporates contextual information from past and future into the multimodal embedding space.
The proposed ResBRNN preserves temporal coherence and increases diversity of the corresponding embeddings.

Second, we propose a Narrator model to generate stories.
Given an input video, the Narrator extracts from it a sequence of important clips.
Then a story is generated by retrieving a sequence of sentences that best matches the clips in the multimodal embedding space,
and concatenating the retrieved sentences.
\fig~\ref{fig:example} shows an example story using the proposed method.
Discovering the important clips is difficult because there is no clear definition as to what visual aspects are important in terms of forming a good story.
To this end,
we formulate the Narrator as a reinforcement learning agent that sequentially observes an input video,
and learns a policy to select clips that maximize the reward.
We formulate the reward as the textual metric between the retrieved story and the reference stories written by human.
By directly optimizing the textual metric,
the Narrator can learn to discover important and diversified clips that form a good story.

Third, 
we have collected a new Video Story dataset (see Section~\ref{sec:dataset} for details) to enable this study, which is publicly available~\footnote{https://zenodo.org/record/2383739}. 
We evaluate our proposed method on this dataset,
and quantitatively and qualitatively show that our method outperforms existing baselines.

The rest of the paper is organized as follows.
Section~\ref{sec:literature} reviews the related work.
Section~\ref{sec:method} delineates the details of the proposed context-aware multimodal embedding and narrator network.
The new Video Story dataset is described in Section~\ref{sec:dataset},
whereas the experimental results and discussion are presented in Section~\ref{sec:experiment}.
Section~\ref{sec:conclusion} concludes the paper.

%% file: table/fig_example.tex
\begin{figure*}[!t]
 \centering
  \begin{minipage}{0.497\textwidth}	\centerline{\includegraphics[width=\linewidth]{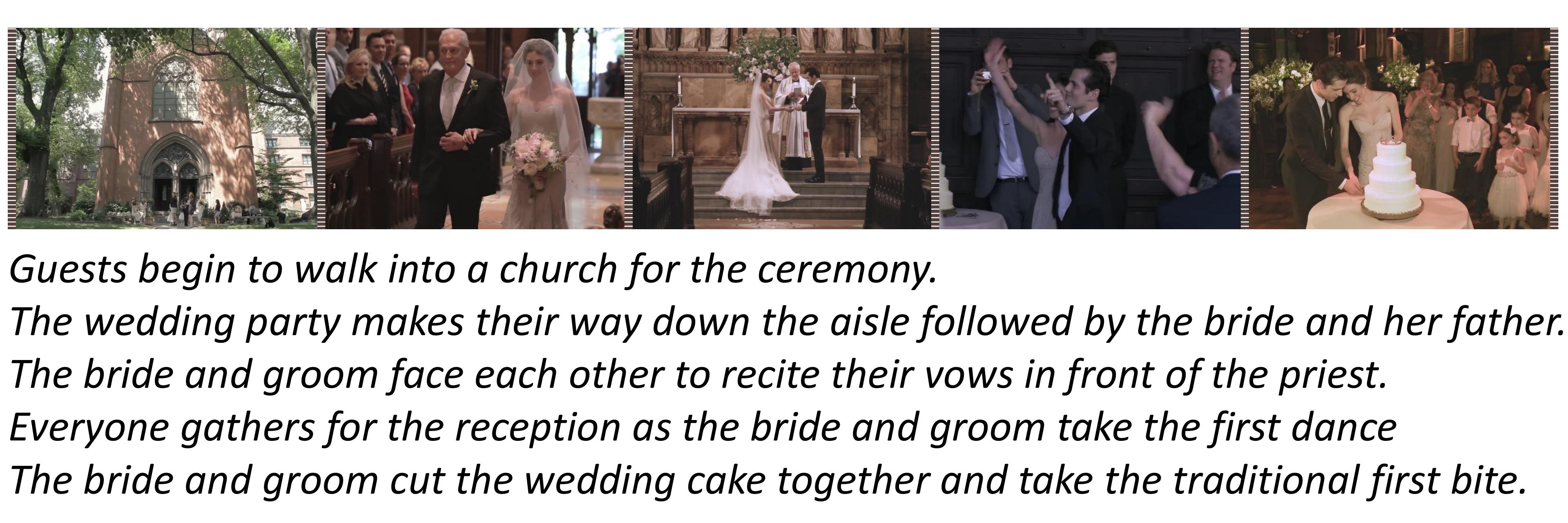}}   \end{minipage}
  \begin{minipage}{0.497\textwidth}	\centerline{\includegraphics[width=\linewidth]{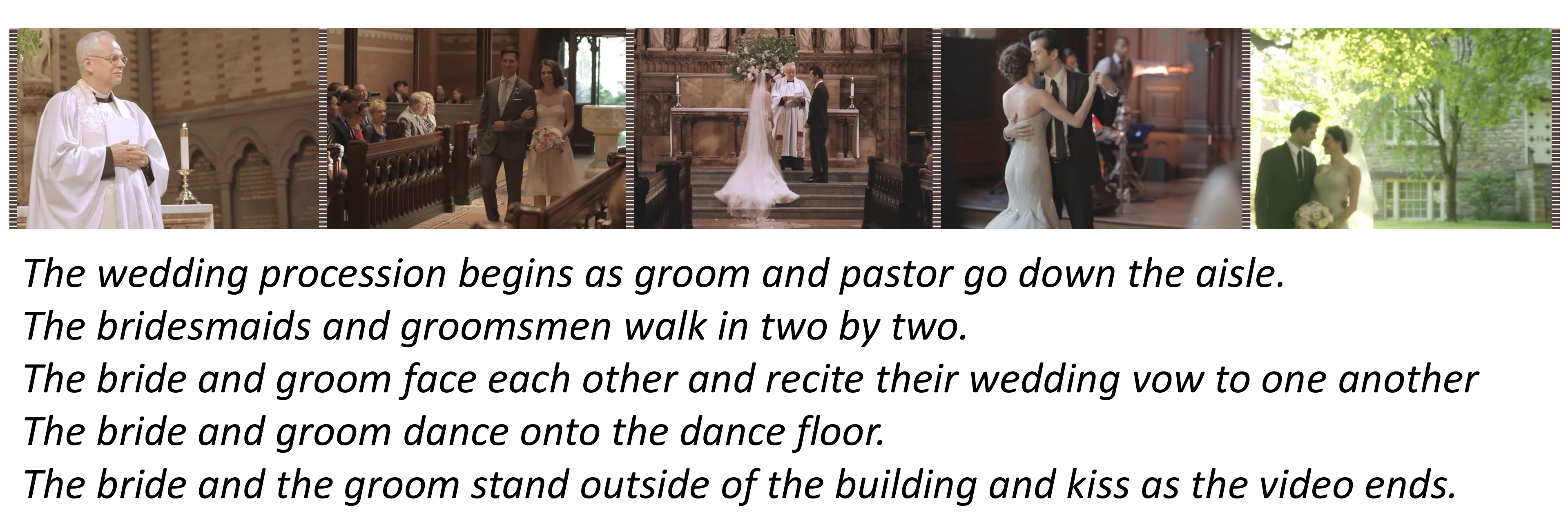}}  \end{minipage}  
  \caption
  	{
  	\small
  		The story written by human (left) and the proposed method (right).
  		We show five sentences and their corresponding key frames uniformly sampled from each story.
  	}	 
  \label{fig:example}
 \end{figure*}  	

%% file: sec_literature.tex
\section{Related Work}
\label{sec:literature}

\subsection{Visual Paragraph Generation}

Bridging vision and language is a longstanding goal in computer vision.
Earlier works mainly target image captioning task,
where a single-sentence factual description is generated for an image~\cite{Mao_CoRR_2014,Xu_ICML_2015}, or a corresponding segment~\cite{Karpathy_CVPR_2015}.
Recent works~\cite{Krause_2017_CVPR,Liang_ICCV_2017} 
aim to provide more comprehensive and fine-grained image descriptions by generating multi-sentence paragraphs.
Generating paragraphs for photo streams have also been explored.
Park and Kim~\cite{Park_NIPS_2015} first studied blog posts,
and propose a method to map a sequence of blog images into a paragraph.
Liu~\etal~\cite{Liu_AAAI_2017} proposes a skip Gated Recurrent Unit (sGRU) to deal with visual variance among photos in a stream.

The progression from single-sentence to paragraph generation also appears in the video captioning task.
Pioneering works~\cite{Rohrbach_GCPR_2014,Yu_CVPR_2016} first studied paragraph generation for cooking videos,
and the more recent work~\cite{Ranjay_ICCV_2017} focuses on dense descriptions of activities.
An encoder-decoder paradigm has been widely studied,
where an encoder first encodes the visual content of a video, followed by a recurrent decoder to generate the sentence.
Existing encoder approach include Convolutional Neural Network (CNN) with attention mechanism~\cite{Junnan_MM,Yao_ICCV_2015,Junnan_ICCV}, mean-pooling~\cite{Venugopalan_HLT_2015}, and RNN~\cite{Ranjay_ICCV_2017,Venugopalan_ICCV_2015,Donahue_PAMI_2017,Junnan_NIPS,Gao_TMM_2018}. 
To capture the temporal structure in more detail, 
hierarchical RNN is proposed specifically for paragraph generation~\cite{Yu_CVPR_2016} where the top level RNN generates hidden vectors that are used to initialize low level RNNs to generate individual sentences.

A common problem in using RNN sentence decoder is that it tends to generate `safe' descriptions that are generic and repetitive. 
This is because the maximum likelihood training objective encourages the use of \textit{n-grams} that frequently appear in training sentences~\cite{Huang_NAACL_2016,Dai_ICCV_2017,Li_HLT_2016}.
In video storytelling task,
this problem can be more severe, 
because a story should naturally contain diverse rather than repetitive sentences.
Another line of work for visual descriptions generation pose the task as a retrieval problem,
where the most compatible description in the training set is transferred to a query~\cite{Park_NIPS_2015,Hodosh_JAIR_2013,Ordonez_NIPS_2011,Socher_TACL_2014,Xu_AAAI_2015,Zhu_ICCV_2015,Dong_TMM_2018}.
Ordonez~\etal~\cite{Ordonez_NIPS_2011} and Park and Kim~\cite{Park_NIPS_2015} select candidate sentences based on visual similarity,
and re-rank them to find the best match.
The other approaches jointly model visual content and text to build a common embedding space for sentence retrieval~\cite{Socher_TACL_2014,Xu_AAAI_2015,Zhu_ICCV_2015,Tsai_ICCV_2017,Faghri_BMVC_2018}.

In this work,
we draw inspiration from both approaches.
We employ CNN+RNN encoders to learn a cross-modality context-aware embedding space,
and generate stories by retrieving sentences.
In this way,
as shown in our experiment,
we are able to generate natural, diverse, and semantically-meaningful stories.

\input{table/fig_embedding}

\subsection{Video Summarization}

Video summarization algorithms aim to aggregate segments of a video that capture its essence.
Earlier works mainly utilize low-level visual features such as interest points, color and motion~\cite{Dufaus_ICIP_2000,Goldman_ACMTG_2006,Lu_TMM_2014}.
Recent works incorporate high-level semantics information such as visual concept~\cite{Song_CVPR_2015}, vision-language semantics~\cite{Plummer_CVPR_2017}, human action~\cite{Varini_TMM_2017,Pablos_TMM_2018}, human-object interaction~\cite{Lu_CVPR_2013,Xu_2019_CVPR} and video category~\cite{Potapov_ECCV_2015,Zhang_CVPR_2016}.
However, there has been no established standard as to what constitutes a good visual summary.
Evaluating the visual similarity between a machine-generated summary and a user-generated summary is difficult, and researchers need to define objective functions to measure the quality of a visual summary.
Vasudevan~\etal~\cite{Vasudevan_MM_2017} select keyframes based on their relevance to the query.
Other researchers have manually defined different criteria such as interestingness, representativeness and uniformity~\cite{Gygli_CVPR_2015,Plummer_CVPR_2017}.
Those criteria suffer from subjectiveness, which also makes evaluation challenging. 

Yeung~\etal~\cite{Yeung_CVPRW_2014} show that textual distance is a better measurement of semantic similarity compared to visual distance.
Therefore, they propose VideoSET where video segments are annotated with text, and evaluation is performed in text domain. 
Sah~\etal~\cite{Sah_WACV_2017} and Chen~\etal~\cite{V2TS} further propose methods to generate a textual summary for a video.
However, both methods generate disconnected captions for individual clips while ignoring the storyline of an event.

In this work, we propose a novel method to select the important video clips, which circumvents the need for pre-defined criteria.
In contrast,
we directly leverage the reference stories written by human for supervision,
and learn a clip selection policy that maximizes the language metric of the generated stories.
The advantage of our method is that we can train with a more well-defined and objective goal.

\subsection{Learning Task-specific Policies}

We draw inspiration from recent works that use REINFORCE~\cite{Williams_ML_1992} to learn task-specific policies.
Ba~\etal~\cite{Ba_CoRR_2014} and Mnih~\etal~\cite{Mnih_2014_NIPS} learn spatial attention policies for image classification, 
whereas other works apply REINFORCE for image captioning~\cite{Xu_ICML_2015,Liu_ICCV_2017,Ranzato_ICLR_2016,Rennie_CVPR_2016}.
In video domain, Yeung~\etal~\cite{Yeung_2016_CVPR} learn a policy for the task of action detection.
Lan~\etal~\cite{Lan_CVPR_2018} propose a reinforcement learning agent for video fast-forwarding.

%% file: table/fig_embedding.tex
\begin{figure*}[!t]
 \centering
  	\includegraphics[width=1.0\textwidth]{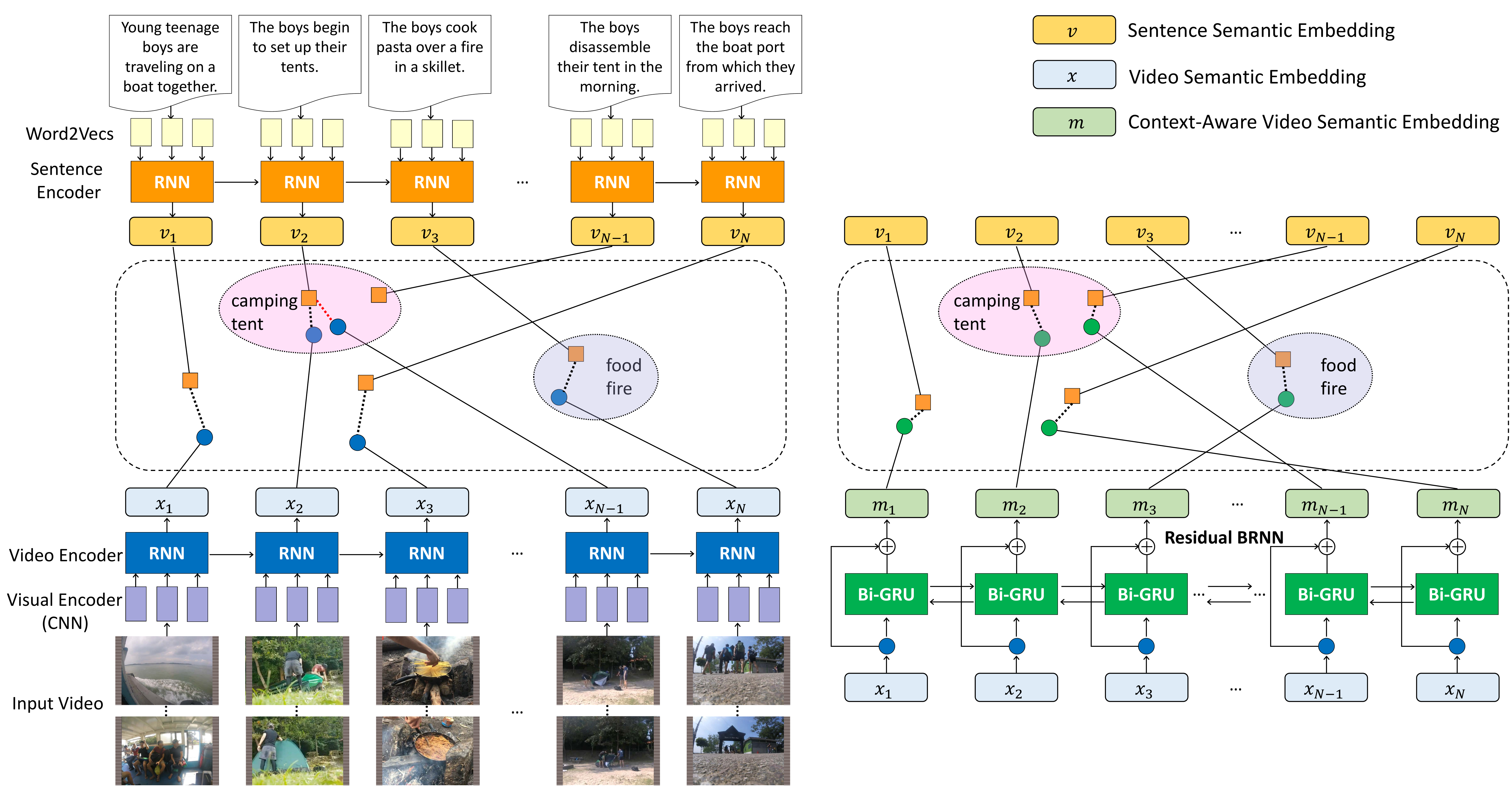}
  \caption
  	{
  	\small
		\textit{Local-to-global} multimodal embedding learning framework. Input is a sequence of video clips and sentences.
		Left side shows local embedding learning,
		where each clip is encoded by a CNN+RNN network to obtain video semantic embedding $\Vec{x}$,
		and each sentence is encoded by a RNN to obtain sentence semantic embedding $\Vec{v}$.
		Right side shows global embedding learning, where the entire sequence of $\Vec{x}$ is fed to the ResBRNN to obtain the context-aware video semantic embedding $\Vec{m}$.
		ResBRNN models the dynamic flow of the entire story, hence the penultimate clip can be correctly mapped to its corresponding sentence.
	  } 
  \label{fig:embedding}
 \end{figure*}

%% file: sec_method.tex
\section{Method}
\label{sec:method}

The proposed video storytelling method involves two sub-tasks:
(a)~discover the important clips from a long video, and
(b)~generate a story for the selected video clips with sentence retrieval.
To address this, the proposed method includes two parts.
In the first part, 
we propose a {\it context-aware multimodal embedding learning framework} where the learned embedding captures the event dynamics.
In the second part, 
we propose a {\it Narrator model} that learns to discover important clips to form a succinct and diverse story.

\subsection{Context-Aware Multimodal Embedding}
\label{sec:method_embedding}

We propose a two-step \textit{local-to-global} multimodal embedding learning framework.
An overview of the framework is shown in \fig~\ref{fig:embedding}.
In the first step, we follow the image-sentence ranking model proposed by~\cite{Kiros_NIPSW_2014} to learn a local clip-sentence embedding.
In the second step, we propose ResBRNN that leverages video-story pairs to model the temporal dynamics of a video,
and incorporates global contextual information into the embedding.
We will first describe the clip-sentence embedding.

\noindent\textbf{Encoder}
The model by~\cite{Kiros_NIPSW_2014} consists of two branches: a Long Short-Term Memory (LSTM) network to encode a sentence into a fixed-length vector,
and a Convolutional Neural Network (CNN) followed by linear mapping to encode an image.
Similarly, we construct two encoders for sentence and video, respectively.
The sentence encoder is a RNN that takes the Word2Vecs~\cite{Mikolov_NIPS_2013} of a word sequence as input, 
and represents the sentence with its last hidden state $\Vec{v}$. 
For the video encoder,
inspired by the encoder-decoder model~\cite{Venugopalan_ICCV_2015,Donahue_PAMI_2017},
we use a RNN that take in the output of a CNN applied to each input frame,
and encodes the video as its last hidden state $\Vec{x}$.
In terms of the recurrent unit, we use Gated Recurrent Unit (GRU)~\cite{Cho_EMNLP_2014} instead of LSTM,
because GRUs have been shown to achieve comparable performance to LSTM on several sequential modeling tasks while being simpler~\cite{Chung_CoRR_2014}.
The GRU uses gating units to modulate the flow of information, 
specified by the following operation:
\begin{equation}
\label{eq:gru}
\begin{split}
	&\Vec{r}_t=\sigma(\Mat{W}_{rx}\Vec{x}_t+\Mat{W}_{rh}\Vec{h}_{t-1}+\Vec{b}_r) \\
	&\Vec{z}_t=\sigma(\Mat{W}_{zx}\Vec{x}_t+\Mat{W}_{zh}\Vec{h}_{t-1}+\Vec{b}_z) \\
	&\Vec{\tilde{h}}=tanh(\Mat{W}_{hx}\Vec{x}_t+\Mat{W}_{hh}\Vec{r}_t\odot\Vec{h}_{t-1}+\Vec{b}_h)\\
	&\Vec{h}_t=\Vec{z}_t\odot\Vec{h}_{t-1}+(1-\Vec{z}_t)\odot\Vec{\tilde{h}}\\
\end{split}	
\end{equation}
where $\odot$ denotes element-wise multiplication, $\sigma$ are Sigmoid functions, $t$ is the current time, $\Vec{x}_t$ is the input, $\Vec{\tilde{h}}$ is the current hidden state, $\Vec{h}_t$ is the output. $\Vec{r}_t$ and $\Vec{z}_t$ are \textit{reset gate} and \textit{update gate}, respectively. 

In this work, the dimensionality of the hidden state of both encoder RNNs are set to be 300.
The Word2Vecs are 300-dimensional vectors.
The CNN we use on video frame is the 101-layer ResNet~\cite{He_CVPR_2016} that outputs 2048-dimensional vectors.
Following~\cite{Venugopalan_ICCV_2015,Venugopalan_HLT_2015},
a video is sampled on every tenth frame.

\noindent\textbf{Clip-Sentence Ranking Loss}
We aim to learn a joint embedding space where distance between embeddings reflects semantic relation.
We assume that paired video clip and sentence share the same semantics,
hence they should be closer in the embedding space.
We define a similarity scoring function $s(\Vec{v},\Vec{x})=\Vec{v}\cdot\Vec{x}$ for sentence embedding $\Vec{v}$ and clip embedding $\Vec{x}$,
where $\Vec{v}$ and $\Vec{x}$ are first scaled to have unit norm (making $s$ equivalent to cosine similarity).

We then minimize the following pairwise ranking loss:
\begin{equation}
\label{eq:rankloss1}
\begin{split}
	\min_{\theta_{en}}\sum_{x}\sum_{k}\max\{0,\alpha-s(\Vec{x},\Vec{v})+s(\Vec{x},\Vec{v}_k)\} \\
	+\lambda\sum_{v}\sum_{k}\max\{0,\alpha-s(\Vec{v},\Vec{x})+s(\Vec{v},\Vec{x}_k)\}
\end{split}	
\end{equation}
where $\theta_{en}$ denote all the parameters to be learned (weights of the two encoder RNNs),
$\Vec{v}_k$ is the negative paired (non-descriptive) sentence for clip embedding $\Vec{x}$,
and vice-versa with $\Vec{x}_k$.
The negative samples are randomly chosen from training set and re-sampled every epoch.
$\alpha$ denotes the margin and is set to 0.1 in our experiment.
The weight $\lambda$ balances the strengths of the two ranking terms.
We found $\lambda=0.5$ produces the best results in our experiment.

\vspace{2mm}   
\noindent\textbf{Temporal Dynamics with Residual Bidirectional RNN}\\
One major challenge for video storytelling is that visually similar video clips could have different semantic meanings depending on the global context.
For example, in \fig~\ref{fig:embedding},
the second and the penultimate clip both show some people around a tent.
However, depending on the sequence of events that happen before and after, their descriptions differ from `set up their tent' to `disassemble their tent'.
It is therefore very important to model the entire flow of the story.
To this end, we propose a ResBRNN model, that builds upon the embedding learned in the first step, and leverages video-story pairs to incorporate global contextual information into the embedding. 

ResBRNN takes a sequence of clip embedding vectors {\small $\Mat{X}=\{\Vec{x}_1,\Vec{x}_2,...,\Vec{x}_N\}$},
and outputs a sequence of context-aware clip embedding vectors {\small $\Mat{M}=\{\Vec{m}_1,\Vec{m}_2,...,\Vec{m}_N\}$} of the same size.
The role of ResBRNN is to refine the embeddings {\small $\Mat{X}$} by incorporating past and future information.
Inspired by the residual mapping~\cite{He_CVPR_2016},
we add a shortcut connection from the input directly to the output.
Intuitively, 
to refine the embedding,
it is much easier to add in fine-scale details via residual {\small $\Mat{M}-\Mat{X}$},
where the shortcut connection enables an identity mapping from {\small $\Mat{X}$} to {\small $\Mat{M}$}.
Compared with regular BRNN, the identity mapping provides a good initialization in the output space that is much more likely to be closer to the optimal solution.
Therefore learning would be more effective.
Our experiment in Section~\ref{sec:experiment} proves that ResBRNN achieves significant performance improvement.
To enable this,
we re-write the GRU formula in (\ref{eq:gru}) into a compact form: {\small$\Vec{h}_t=GRU(\Vec{x}_t,\Vec{h}_{t-1};\Mat{W})$},
and define the operation of the proposed ResBRNN as:
\begin{equation}
\begin{split}
\Vec{h}_t^f &= GRU(\Vec{x}_t,\Vec{h}_{t-1}^f;\Mat{W}) \\
\Vec{h}_t^b &= GRU(\Vec{x}_t,\Vec{h}_{t-1}^b;\Mat{W})\\
\Vec{m}_t&=\Vec{x}_t+\Vec{h}_t^f+\Vec{h}_t^b
\end{split}	
\end{equation}
where $f$ denotes the forward pass and $b$ denotes the backward pass.
The parameters $\Mat{W}$ for the two passes are shared to reduce the number of parameters.

During training, we minimize a video-story pairwise ranking loss:
\begin{equation}
\label{eq:rankloss2}
\small
\begin{split}
\min_{\theta}\sum_{m}\sum_{k}\max\{0,\beta-s(\Vec{m},\Vec{v})+s(\Vec{m},\Vec{v}_k)\} \\
+\lambda\sum_{v}\sum_{m}\max\{0,\beta-s(\Vec{v},\Vec{m})+s(\Vec{v},\Vec{m}_k)\}
\end{split}	
\end{equation}
$\Vec{v}$ and $\Vec{m}$ are paired sentence and clip embeddings from the same story (or video), 
while $\Vec{v}_k$ (or $\Vec{m}_k$) are the negative paired sample to $\Vec{m}$ (or $\Vec{v}$).
$\theta$ include the parameters of ResBRNN and the parameters of the encoders $\theta_{en}$.
$\beta$ denotes the margin. 
We set this margin to be larger than $\alpha$ ($\beta=0.2$) to apply harder constraints on the embeddings.

\noindent\textbf{Optimization}
We first optimize the clip-sentence ranking loss (\ref{eq:rankloss1}) until validation error stops decreasing.
Then we optimize the video-story ranking loss (\ref{eq:rankloss2}).
We use ADAM~\cite{Kingma_CoRR_2014} optimizer with a first momentum coefficient of 0.8 and a second momentum coefficient of 0.999.

\input{table/fig_narrator}

\subsection{Narrator Network}
The goal of the Narrator is to take in a long video and output a sequence of important video clips that form a story.
The model is formulated as a reinforcement learning agent that interacts with a video over time.
The video is first processed by the video encoder CNN+RNN that has been trained following the method described in Section~\ref{sec:method_embedding}.
The video encoder produces a sequence of features {\small $\Mat{X}=\{\Vec{x}_1,\Vec{x}_2,...,\Vec{x}_N\}$} (300-d hidden state) for $N$ frames, 
where the frames are sampled on every tenth frame from the video. 
At timestep $n$, $\Vec{x}_n$ contains semantic information of the current and all previous frames.
The Narrator sequentially observes $\Mat{X}$,
and decides both when to sample a clip and the length of the sampled clip.
An overview of the model is shown in \fig~\ref{fig:narrator}.
We now describe the Narrator in detail.

\noindent\textbf{Candidate Gate}
At each timestep $n$, a binary gate $R_n$ is used to decide whether the current position is a candidate to sample a clip.
The gate is defined as
\begin{equation}
R_n(\Vec{x}_n,\Vec{x}_p)=
\begin{cases}
1 \hspace{2em}\text{ if } s(\Vec{x}_n,\Vec{x}_p)<\tau\\
0 \hspace{2em}\text{ otherwise}
\end{cases}
\end{equation}
where $s(\Vec{x}_n,\Vec{x}_p)=\Vec{x}_n\cdot\Vec{x}_p$ is the similarity score for the normalized frame embeddings of the current position $n$ and the previous sample position $p$. 
The candidate gate serves as an attention mechanism. 
It reject the current frame if it is semantically similar with the previous sampled frame.
It enforces succinctness and diversity to the story, 
and also makes training easier by reducing the size of the agent's possible action space.
In this work, we set the threshold $\tau$ to be 0.7.

\noindent\textbf{Clip Indicator}
If $R_n=1$ at timestep $n$,
the current frame embedding $\Vec{x}_n$ is processed by a function $f_c(\Vec{x}_n;\theta_c)$ to get the clip indicator $c_n$, 
which is a binary value that signals whether a clip should be sampled at the current position.
$f_c(\Vec{x}_n;\theta_c)$ is a function learned from data that decides how important the current frame is for a story.
During training,
$c_n$ is sampled from a Bernoulli distribution parameterized by $f_c(\Vec{x}_n;\theta_c)$ defined as
\begin{equation}
	f_c(\Vec{x}_n;\theta_c)=\max(0,\sigma(\Mat{W}_c\Vec{x}_n)+b_c)
\end{equation} 
where $\theta_c=\{\Mat{W}_c,b_c\}$, $\sigma$ denotes the Sigmoid function,
$\Mat{W}_c$ denotes the weights for a fully-connected layer,
and $b_c$ is a scalar value to offset the probability distribution.
Since our goal is to generate succinct rather than dense descriptions,
the clips should be sampled sparsely from the video.
Hence we initialize $b_c$ to be $-0.4$, so that $f_c(\Vec{x}_n;\theta_c)$ is more biased towards zero.

At test time, $c_n$ is computed as
\begin{equation}
c_n=
\begin{cases}
1 \hspace{2em}\text{ if } f_c(\Vec{x}_n;\theta_c)>\epsilon\\
0 \hspace{2em}\text{ otherwise}
\end{cases}
\end{equation}
where $\epsilon$ is a threshold that controls the succinctness of the story. 
In our experiment, we set $\epsilon=0.2$ based on performance on the validation set.

\noindent\textbf{Clip Length}
If $c_n$ is set to 1, the agent will sample a clip centered at the current position with length $l_n$ that refers to the number of frames
(note that the frames are sampled on every tenth frame from the video).
$l_n$ is computed by a function $f_l(\Vec{x}_n;\theta_l)=\kappa\cdot\sigma(\Mat{W}_l\Vec{x}_n+\Vec{b}_l)$,
where $\kappa$ is a scaling constant and is set to 40.
During training, $l_n$ is stochastically sampled from a Gaussian distribution with a mean of $f_l(\Vec{x}_n;\theta_l)$ and a fixed variance.
At test time, $l_n=f_l(\Vec{x}_n;\theta_l)$.

\noindent\textbf{Storytelling}
As shown in \fig~\ref{fig:sub:narrator2}, we take all the clips sampled from a video and compute their 
context-aware semantic embeddings $m$ following the method in Section~\ref{sec:method_embedding}.
Meanwhile, we compute the sentence semantic embeddings $v$ for all candidate sentences.
Then we generate the story by retrieving a set of sentences that best match the sampled clips in the embedding space.
Note that the candidate pool consists of all sentences from the training set.

\input{table/tbl_statistic}

\subsection{Narrator Training}

There are two challenges for learning effective clip selection policy.
First,
it is difficult to visually evaluate the clip proposals~\cite{Yeung_CVPRW_2014}.
Second, 
the sentences retrieved from training set can introduce bias to the story.
To address those challenges,
we learn the clip selection policy by directly optimizing the generated story in text domain.
Inspired by~\cite{Ranzato_ICLR_2016,Rennie_CVPR_2016},
we evaluate the story with NLP metrics such as BLEU~\cite{Papineni_ACL_2002}, METEOR~\cite{Banerjee_ACL_2005} or CIDEr~\cite{Vedantam_CVPR_2015}.
Optimizing on NLP metrics has two advantages:
(1) it is more practical, well-defined and
semantically-meaningful compared with optimizing on the visual content.
(2) the Narrator takes account of the bias from the candidate sentences. 

The training of Narrator is a non-differentiable process because
(1) the NLP metrics are discrete and non-differentiable,
and (2) the clip indicator and clip length outputs are non-differentiable components.
We take advantage of the REINFORCE algorithm~\cite{Williams_ML_1992} that enables learning of non-differentiable models. 
We first briefly describe REINFORCE below.
Then we introduce a variance-reduced reward to learn effective clip selection policy.

\noindent\textbf{REINFORCE}
Given a space of possible action sequences $A$,
the policy of the agent, in our case $f_c$ and $f_l$, \mbox{induces} a distribution $p_\theta(a)$ over $a\in A$ parameterized by \mbox{\small $\theta=\{\theta_c,\theta_l\}$}.
The objective of REINFORCE is to maximize the expected reward $J(\theta)$ defined as
\begin{equation}
	J(\theta)=\sum_{a\in A}p_\theta(a)r(a)
\end{equation}
where $r(a)$ is a reward assigned to each individual action sequence.
The gradient of the objective is
\begin{equation}
\nabla J(\theta)=\sum_{a\in A}p_\theta(a)\nabla \log p_\theta(a)r(a)
\end{equation}

Maximizing $J$ is non-trivial due to the high-dimensional space of possible action sequences.
REINFORCE addresses this by approximating the gradient equation with Monte Carlo sampling:
\begin{equation}
	\nabla J(\theta)\approx \frac{1}{M} \sum_{i=1}^{M} \sum_{n=1}^{N}\nabla \log \pi_\theta(a_n^i|x_{1:n}^i,a_{1:n-1}^i) R^i
	\label{eq:reinforce}
\end{equation}
where $\pi_\theta$ is the agent's current policy. 
At time step $n$, $a_n$ is the policy's current action (\ie~clip indicator $c_n$ and clip length $l_n$),
$x_{1:n}$ are the past states of the environment including the current (frame embeddings),
and $a_{1:n-1}$ are the past actions.
$R^i$ is the reward received by running current action sequence to the end. 
The approximate gradient is computed by running the agent's current policy for $M$ episodes.

In this work, we use the CIDEr score for the generated story as the reward.
Empirically, we find that optimizing over CIDEr achieves the best overall performance compared with other metrics,
which has also been observed by \cite{Liu_ICCV_2017,Rennie_CVPR_2016}.

\noindent\textbf{Variance-reduced reward}
The gradient estimate in (\ref{eq:reinforce}) may have high variance.
Therefore it is common to use a baseline reward $b$ to reduce the variance, and so the gradient equation becomes:
\begin{equation}
	\nabla J(\theta)\approx \frac{1}{M} \sum_{i=1}^{M} \sum_{n=1}^{N}\nabla \log \pi_\theta(a_n^i|x_{1:n}^i,a_{1:n-1}^i) (R^i-b)
\end{equation}

The baseline reward is often estimated with a separate network~\cite{Mnih_2014_NIPS,Liu_ICCV_2017,Ranzato_ICLR_2016,Yeung_2016_CVPR}.
For our model, there are two drawbacks of using a separate network for baseline estimation.
First, it introduces extra parameters to be learned.
Second, a baseline estimator that only operates on the visual input $x$ does not account for the bias introduced by the candidate sentences. 
Therefore, we propose a greedy approach that computes a different baseline for each video.
For each video, we first randomly sample a set of clips,
and retrieval a story using those clips.
Then we repeat the above procedure $K$ times, and take the average of the $K$ stories' CIDEr score as the baseline reward for that video ($K=10$). 
In this way, a policy receives a positive reward if it performs better than a random policy, and negative otherwise.

%% file: table/fig_narrator.tex
\begin{figure*}[!t]
  
  \begin{subfigure}{1\linewidth}
  	\centering
		\includegraphics[width=0.92\linewidth]{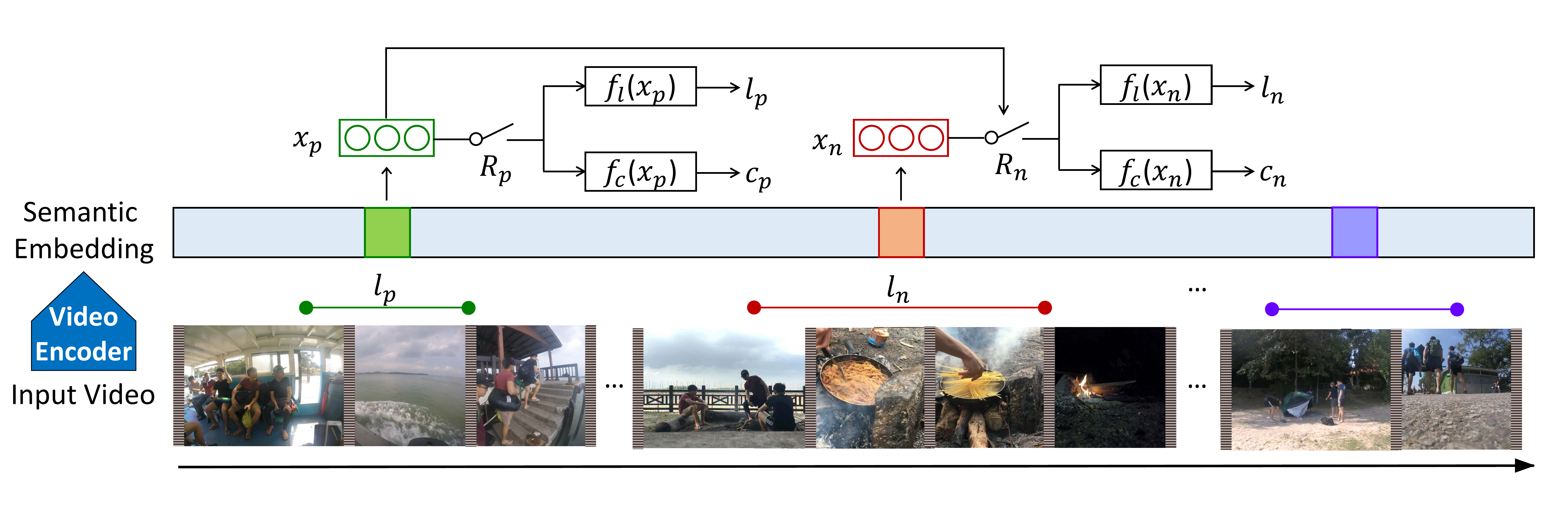}
		\vspace{-3ex}
		\caption{ }
		\label{fig:sub:narrator1}
  \end{subfigure}
  \begin{subfigure}{1\linewidth}
  	\centering
  	\includegraphics[width=0.92\linewidth]{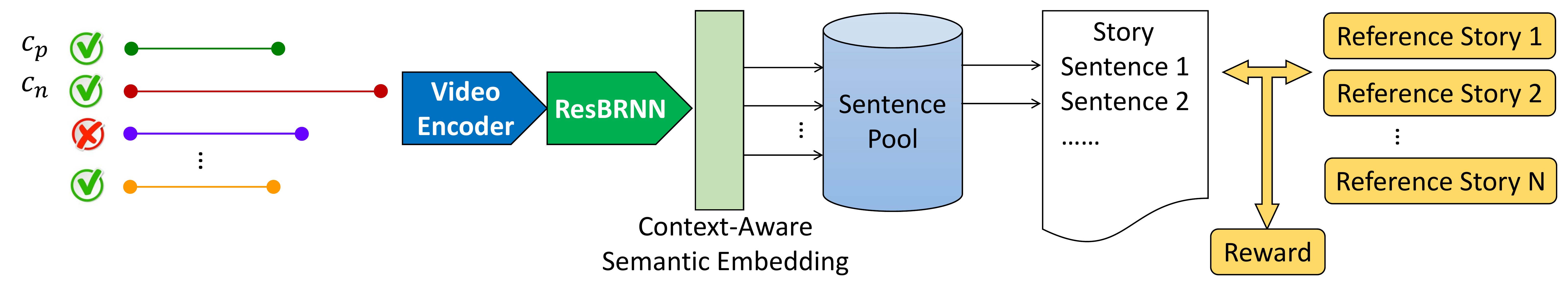}
  	\vspace{-1ex}
  	\caption{ }
  	\label{fig:sub:narrator2}
  \end{subfigure}
  \caption
  	{
  	\small
		(a) Narrator Network. Input is a sequence of encoded video frame features and output is a sequence of clip proposals. 
		Here, we illustrate an example of forward pass. At timestep $n$, the agent observes the frame feature $\Vec{x}_n$ (\textcolor{red}{red}), and decides the current position is a candidate ($R_n=1$) because $\Vec{x}_n$ is semantically different from the previous sampled frame (\textcolor{green}{green}).
	Then the agent produces a clip indicator $c_n$ that indicates whether the current frame is important.
	If $c_n=1$, a clip with length $l_n$ is sampled at the current position.
	The agent continues proceeding until it reaches the end of the video.
	(b) A story is generated by retrieving sentences for the sampled clips. The textual metric of the story is computed as the reward to train the Narrator.
  	} 
  \label{fig:narrator}
 \end{figure*}

%% file: table/tbl_statistic.tex
\begin{table*}[!t]
	\centering
	\caption
		{
		\small	
		Comparison of Video Story with existing video description\&visual story datasets. 
		Note that SIND~\cite{Huang_NAACL_2016} is an image dataset, and * marks the number of albums
		}
	\label{tbl:statistic}
	\begin{tabular}{l|c|c|c|c|c|c} 
		\toprule
		Dataset & Domain &Num. videos & Avg. video length &Avg. text length& Avg. sent length &Vocabulary \\
		\midrule
		MSR-VTT~\cite{Xu_CVPR_2016} & open &7180 & 20.65s & \hspace{8pt}9.6 & \hspace{4pt}9.6 & 24,549 \\
		TaCos M-L~\cite{Rohrbach_GCPR_2014} & cooking &185 & 5m 5s & 115.9 & \hspace{4pt}8.2 & \hspace{4pt}2,833 \\
		ActivityNet Captions~\cite{Ranjay_ICCV_2017} & open & 20k & 180s & \hspace{4pt}52.5 & 13.5 & - \\
		SIND~\cite{Huang_NAACL_2016} & Flickr album&10,117* & 5 images & \hspace{4pt}51.0 & 10.2 & 18,200\\		
		VideoSet~\cite{Yeung_CVPRW_2014} & TV \& egocentric &11 & 3h 41m& 376.2 & \hspace{4pt}8.6& \hspace{4pt}1,195\\
		\midrule
		Video Story& open & 105 & 12m 35s & 162.6 & 12.1 & \hspace{4pt}4,045\\
		\bottomrule
	\end{tabular}
\end{table*}

%% file: sec_dataset.tex
\section{Dataset}
\label{sec:dataset}

Since video storytelling is a new problem,
we have collected a Video Story dataset to enable research in this direction.
We choose four types of common and complex events (\ie~birthday, camping, Christmas and wedding),
and use keyword search to download videos from Youtube.
Then we manually selected 105 videos with sufficient inter-event and intra-event variation.
The stories are collected via Amazon Mechanical Turk.
For each video we asked crowd workers to write their stories following three rules:
(1) The story should have at least 8 sentences.
(2) Each sentence should have at least 6 words.
(3) The story should be coherent and relevant to the event.
We then asked workers to label the
start and end time in the video where each sentence occurred.
Each video is annotated by at least 5 different workers.
In total, we collected 529 stories.

Table~\ref{tbl:statistic} shows the statistics of the Video Story dataset in comparison with existing datasets.
Compared with video captioning datasets (\ie~MSR-VTT~\cite{Xu_CVPR_2016}, \mbox{TaCos M-L} \cite{Rohrbach_GCPR_2014}, ActivityNet Captions~\cite{Ranjay_ICCV_2017}) and album story dataset (\ie~SIND~\cite{Huang_NAACL_2016}),
Video Story dataset has longer videos (12 min 35 sec in average) and longer descriptions (162.6 words in average).
Moreover, the sentences in Video Story are more sparsely distributed across the video (55.77 sec per sentence).
Compared with VideoSet~\cite{Yeung_CVPRW_2014} for video summarization study that only has 11 videos,
Video Story dataset has more videos in open domain,
the sentences are more diverse,
and the vocabulary size is also larger.

%% file: sec_experiment.tex
\section{Experiment}
\label{sec:experiment}

We compare the proposed approach with state-of-the-art methods and variations of the proposed model using quantitative measures and user study.
%
Experiments are performed on the Video Story dataset.
We randomly split 70\% (73 videos) as training set, 15\% (16 videos) as validation set and the others (16 videos with 4 per event category) as test set.
We perform different tasks to validate the proposed multimodal embedding learning framework and the Narrator model.

\subsection{Multimodal Embedding Evaluation}

\subsubsection{Sentence retrieval task}
The goal of this task is to retrieve a sequence of sentences given a sequence of clips as query.
We consider a withheld test set,
and evaluate the median rank of the closest ground truth (GT) sentence and Recall@K, which measures the fraction of times a GT sentence was found among the top K results.
Since the temporal location of clips are sparse in the Video Story dataset,
overlapping clips usually share similar description.
Therefore for a clip, we also include the paired sentences for its overlapping clips into the GT sentences.
In average, each clip has 4.44 GT sentences.

In this work, we compare with the following baseline methods to fully evaluate our model.
\begin{itemize}
	\item
	\textbf{Random:} A naive baseline where sentences are randomly ranked. 
   \item
   \textbf{Category Random:} Only sentences from the same category as the test clip are considered candidates in the random ranking.
   \item   
  \textbf{CCA:}
	Canonical Correlation Analysis (CCA) has been used to build an image-text space~\cite{Socher_CVPR_2010}.
	We use CCA to learn a joint video-sentence space.
	The video feature is the average of frame-level ResNet outputs,
	and the sentence is represented by the average of Word2Vecs.
	Then we project clips and sentences into the joint space for ranking.
	\item
	\textbf{m-RNN~\cite{Mao_CoRR_2014}:}
	m-RNN consists of two sub-networks that encode sentence and image respectively, which interact with each other in a multimodal layer.
	To apply m-RNN for video data, 
	we replace the original CNN image encoder with mean-pooling of frame-level ResNet outputs.
	\item
	\textbf{Xu~\etal~\cite{Xu_AAAI_2015}:}
	This is a sentence-video joint embedding model,
	where sentences are embedded with a dependency-tree structure model.
	We replace the original CNN with ResNet.
	\item
	\textbf{EMB:}
	This is the first local step of our embedding learning framework.
	Sentences are ranked based on their distance from the video in the embedding space.
	Note that EMB is similar to the image-sentence ranking model in~\cite{Liu_AAAI_2017,Kiros_NIPSW_2014} and the movie-text ranking model in~\cite{Zhu_ICCV_2015}.
	\item
	\textbf{BRNN:}
	We use a BRNN in the second step to model the context,
	which makes it a variant to our model without the residual mapping.
	\item
	\textbf{ResBRNN:}
	The proposed context-aware multimodal embedding learning framework.
\end{itemize}

The results are shown in \tab~\ref{tbl:retrieval}.
Our local embedding learning method (EMB) is a strong baseline that outperforms existing methods.
The proposed context-aware embedding learning framework (ResBRNN) further improves the performance by a large margin.
The improvement 
suggests that by incorporating information from contextual clips, 
the model can learn temporal coherent embeddings that are more representative of the underlying storyline.
BRNN without residual mapping performs worse than ResBRNN and EMB,
which demonstrates the importance of residual mapping.

\input{table/tbl_retrieval}
\input{table/tbl_story}
\input{table/fig_full_example_1}
\vspace{2mm}
\subsubsection{Story generation task}
In this task,
we generate stories for test videos with different methods.
In order to evaluate our embedding learning framework,
we need to first fix on a good sequence of clips to represent the story visually.
We employ a `Pseudo-GT' clip selection scheme.
For each test video,
we compare a human annotator's selected clips with other annotators',
and choose the sequence of clips that have the largest overlap with others,
where overlap is computed as intersection-over-union (IoU).
For quantitative measures, we compare the generated story with reference stories and compute NLP metrics of language similarity
(\ie~BLEU-N \cite{Papineni_ACL_2002}, ROUGE-L~\cite{Lin_ACL_2004}, METEOR~\cite{Banerjee_ACL_2005} and CIDEr~\cite{Vedantam_CVPR_2015}) using MSCOCO evaluation code~\cite{Chen_CoRR_2015}.
We would like to first clarify that each of the above metrics has its own strengths and weaknesses~\cite{Kilickaya_EACL_2017}.
For example, BLEU metric computes an \textit{n-gram} based precision,
thus it favors short and generic descriptions, and may not be a good measure for fully evaluating the semantic similarity of long paragraphs.
CIDEr and METEOR have shown reasonable correlation with human judgments in image captioning~\cite{Vedantam_CVPR_2015,Kilickaya_EACL_2017}. 
Therefore, we consider those two metrics as primary indicators of a model's performance. 
Nonetheless, we report results for all metrics.

\input{table/tbl_narrator}

For this task, 
we compare against the baselines in \textit{sentence retrieval} task as well as two state-of-the-art generative models for video captioning and video paragraph generation.
The additional baselines are as follow.
\begin{itemize}
  \item 
  \textbf{S2VT~\cite{Venugopalan_ICCV_2015}}:
	This model encodes each clip with a CNN+RNN model and decodes the sentence with another RNN.
	\item 
	\textbf{H-RNN~\cite{Yu_CVPR_2016}:}
	This model is designed for generating multiple sentences,
	where it uses previous sentences to initialize the hidden state for next sentence generation.
	\item 
	\textbf{ResBRNN-kNN:}
	ResBRNN retrieves the best sentence for each clip in a sequence,
	which may result in duplicate sentences.
	To increase diversity,
	we apply $k$-nearest search.
	For each clip, we first find its $k$-nearest sentences.
	Then we consider the entire sequence, 
	and find the best sequence of \textit{non-duplicate} sentences that have the minimum \textit{total} distance with the clips.
	We use $k=4$ in our experiment.
\end{itemize}

%
%

\tab~\ref{tbl:story} shows the results for story generation with fixed clip selection.
Retrieval based methods generally outperform generative methods,
because generative methods tend to give short, repetitive and low-diversity descriptions.
Among the retrieval methods, 
we observe similar trend of performance improvement as in \tab~\ref{tbl:retrieval}.
ResBRNN-kNN further improves upon ResBRNN by increasing story diversity.
H-RNN has higher BLEU-3 and BLEU-4 score because the BLEU metric favors generic descriptions.
Qualitative examples are shown in \fig~\ref{fig:full_example1}.



\subsection{Narrator Evaluation}

Here we evaluate the Narrator model for clip selection.
Since our end goal is video storytelling, we directly evaluate the NLP metrics of the story retrieved from different clip proposals,
which is a more well-defined and objective evaluation compared with evaluating the overlap of visual content.
We fix the story retrieval method with ResBRNN-kNN,
and compare Narrator with multiple baseline methods for clip selection,
as described below.
\begin{itemize}
  \item 
  \textbf{Uniform:}
	As a naive baseline, we uniformly sample 12 clips per test video, while 12 approximates the average number of sentences per story.
	\item 
	\textbf{SeqDPP~\cite{Gong_NIPS_2014}:}
	A probabilistic model for diverse subshot selection that learns from human-created summaries.
	\item 
	\textbf{Submodular~\cite{Gygli_CVPR_2015}:}
	A supervised approach for video summarization,
	which jointly optimizes for multiple objectives including \textit{Interestingness, Representativeness} and \textit{Uniformity}.
	\item 
	\textbf{vsLSTM~\cite{Zhang_ECCV_2016}:}
	vsLSTM is a bi-directional LSTM network proposed for video summarization.
	The inputs are frame-level CNN features, in our case ResNet outputs.
	The outputs are frame-level importance scores,
	which can be used to select clips following the approach in~\cite{Zhang_ECCV_2016}.
	We compute the GT frame-level importance scores using the same approach as~\cite{Zhang_ECCV_2016},
	where a frame is considered more important if it has been selected by more human annotators.
\end{itemize}

%
%
%

In addition to the above baseline methods, 
we evaluate several variants of the proposed Narrator as ablation study to understand the efficacy of each components.
\begin{itemize}
  \item 
	\textbf{Narrator IoU:}
	Instead of using the CIDEr score of the story as reward, we compute the overlap (IoU) between the Narrator's clip proposal and the annotators' selected clips as reward.
	\item 
	\textbf{Narrator w/o $\Vec{x}$:}
	A variant of the proposed method.
	Inputs are frame-level CNN features instead of the semantic embedding $\Vec{x}$.
	\item 
	\textbf{Narrator w/o $f_l$:}
	A variant of the proposed method without the component $f_l$ to decide clip length.
	Each clip has a fixed length of 20.
\end{itemize}

%
%
 
\tab~\ref{tbl:narrator} shows the evaluation of the retrieved story with various clip selection methods.
The proposed Narrator model outperforms all other methods.
The improvement over Narrator w/o $\Vec{x}$ indicates the representative power of the video semantic embedding,
and the improvement over Narrator w/o $f_l$ suggests that it is useful to have a parameterized clip length compared with fixed.
The state-of-the-art video summarization methods (\ie~Submodular~\cite{Gygli_CVPR_2015}, SeqDPP~\cite{Gong_NIPS_2014} and vsLSTM~\cite{Zhang_ECCV_2016}) only perform slightly better than uniform sampling.
This is because the human-selected clips among different annotators have high variance,
which makes frame-level supervision a weak supervision to learn from.
On the other hand, the stories written by annotators are more congruent,
because different set of clips can still convey similar stories,
thus it is stronger supervision.
For the same reason, using IoU as reward for the Narrator does not perform well.

Note that the performance gaps between clip selection methods are relatively small compared with that of \tab~\ref{tbl:story}.
This is because the ResBRNN-kNN method used in this task can already retrieve relevant and meaningful sentences for a video, 
hence the clip selection methods would make more subtle improvements on the quality of the story. 


\subsection{User Study}

We perform user studies using Amazon Mechanical Turk to observe general users' preference on stories generated by four methods:
(a) the proposed ResBRNN-kNN with Narrator (ResNarrator),
(b) EMB with Pseudo-GT clips  ,
(c) H-RNN~\cite{Yu_CVPR_2016} with Pseudo-GT clips and
(d) GT story.
The results are shown in \tab~\ref{tbl:userstudy}.
Please see \fig~\ref{fig:full_example1} for qualitative examples of the stories.

\input{table/tbl_userstudy}

First, we compare stories generated by the ResNarrator with ones generated by the two baseline methods EMB and H-RNN.
For each of the 16 videos in the test set, we ask 10 users to watch the video,
read the three stories presented in random order,
and rank the stories based on how well they describe the video.
On average, 80.0\% users prefer ResNarrator over EMB, and 88.8\% prefer ResNarrator over H-RNN.
which validate that the proposed method generate better stories compared with the baselines.
 
Then, we ask the user to do a pairwise selection between stories from ResNarrator and GT.
38.1\% users prefer stories generated by ResNarrator over GT stories, which further shows its efficacy.

%% file: table/tbl_retrieval.tex
\begin{table}[!t]
	\centering
	\caption
		{
		\small	
			Sentence retrieval results. \textbf{R@K} is Recall@K (higher is better). \textbf{Medr} is the median rank (lower is better)
		}
	\label{tbl:retrieval}
	\begin{tabular}{l|c c c c} 
		\toprule	
		Method\hspace{6ex} & \bf{R@1} &\bf{R@5} & \bf{R@10} & \bf{Medr}\\
		\midrule
		Random & 0.31 & \hspace{4pt}1.38 & \hspace{4pt}3.98 & 215 \\
		Category Random & 1.18 & \hspace{4pt}5.47 & 16.03 & \hspace{4pt}57 \\		
		CCA & 2.37 & 12.83&23.25 & \hspace{4pt}46 \\
		Xu~\etal~\cite{Xu_AAAI_2015}& 4.72 & 19.85 & 35.46 & \hspace{4pt}31 \\
		m-RNN~\cite{Mao_CoRR_2014} & 5.34 & 21.23 & 39.02 & \hspace{4pt}29 \\	
		\midrule
		EMB & 5.50 & 22.02 & 40.52 & \hspace{4pt}27 \\
		BRNN &5.50 & 20.26 & 36.39 & \hspace{4pt}29 \\ 
		ResBRNN & \bf{7.44} & \bf{25.77} & \bf{46.41} & \hspace{4pt}\bf{22} \\ 
		\bottomrule
	\end{tabular}
\end{table}	

%% file: table/tbl_story.tex
\begin{table*}[!t]
	\centering
	\caption
		{
		\small	
			Evaluation of story generation on Video Story test set with fixed Pseudo-GT clips. 
		}
	\label{tbl:story}
	\vspace{-1ex}
	\begin{center}
	\begin{tabular}{l| c | c c c c c c c} 
		\toprule
		Method\hspace{12ex} & Model Type & ~CIDEr~ & ~METEOR~ & ~ROUGE-L~ & ~BLEU-1~ & ~BLEU-2~ & ~BLEU-3~ & ~BLEU-4~ \\
		\midrule
		S2VT*~\cite{Venugopalan_ICCV_2015} & generative & 64.0&14.3 &28.6&63.3 & 40.6&24.6&15.4 \\
		H-RNN*~\cite{Yu_CVPR_2016}         & generative & 64.6&15.5 &28.8&61.6&41.4 &\bf{26.3} &\bf{16.1}  \\
		\midrule
		Random                             & retrieval  & 30.2 &13.1&21.4&43.1&23.1 & 10.0& 4.8\\		
		CCA                                & retrieval  & 71.8 &16.5 &26.7 & 60.1 & 34.7& 11.8&10.1 \\
		Xu~\etal~\cite{Xu_AAAI_2015}       & retrieval  & 79.5 & 17.7 & 28.0&61.7 & 36.4 & 20.2 & 11.5  \\
		m-RNN~\cite{Mao_CoRR_2014}         & retrieval  & 81.3 & 18.0 & 28.5& 61.9 & 37.0 & 21.1 & 11.8  \\
		\midrule
		EMB                                & retrieval  & 88.8 & 19.1 & 28.9& 64.5 & 39.3 & 22.7 & 13.4  \\
		BRNN                               & retrieval  & 81.0 &18.1 &28.3&61.4 &36.6 &20.3 &11.3 \\
		ResBRNN                            & retrieval  & 94.3 & 19.6 & 29.7&66.0& 41.7& 24.3&14.7\\ 
		ResBRNN-kNN                        & retrieval  & \bf{103.6} &\bf{20.1} &\bf{29.9} &\bf{69.1} &\bf{43.5}&26.1 &15.6  \\ 
		\bottomrule
	\end{tabular}
	\end{center}
	\vspace{-1ex}
\end{table*}	

%% file: table/fig_full_example_1.tex
\begin{figure*}[!t]
  \centering
  \begin{minipage}{1.0\textwidth}
		\centerline{\includegraphics[width=\linewidth]{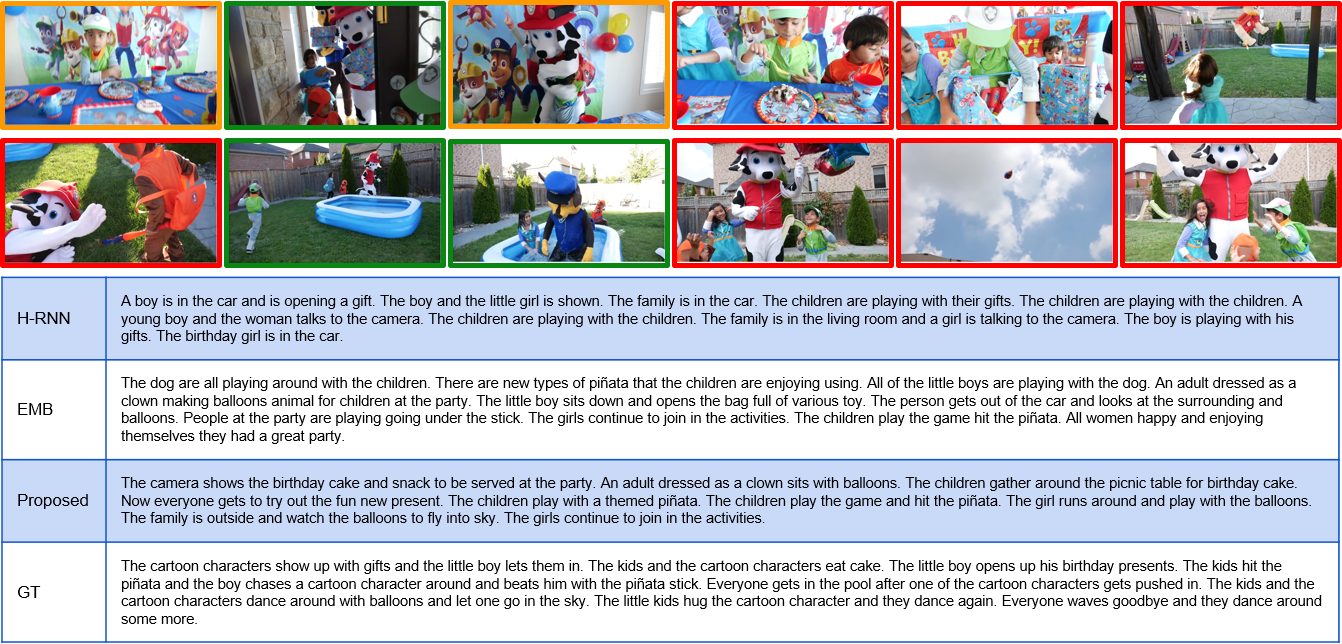}}
		\vspace{1ex}
		\centerline{\includegraphics[width=\linewidth]{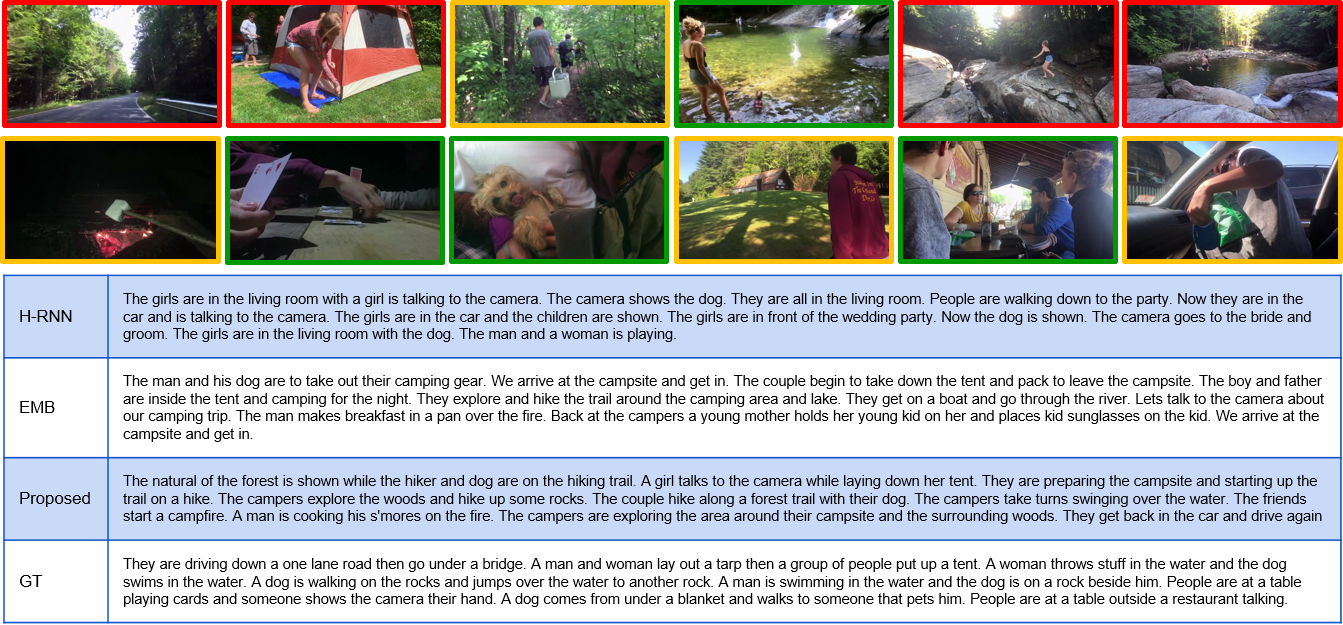}}
  \end{minipage}
  \caption
	{
	\small
	Example stories generated by the proposed method (ResBRNN-kNN+Narrator), two baselines (H-RNN and EMD), and a ground truth.
	The frames are handpicked from the set of clips proposed by Narrator and the ground truth example.
	\textcolor{ForestGreen}{Green} boxes highlight the frames from GT,
	\textcolor{orange}{orange} boxes highlight the frames from Narrator proposals,
	while \textcolor{Red}{red} boxes highlight frames shared by both. 
	}
  \label{fig:full_example1}
 \end{figure*}  	

%% file: table/tbl_narrator.tex
\begin{table*}[!t]
	\caption
		{
		\small	
			Evaluation of story generation on Video Story test set with different clip selection methods and fixed story retrieval method
		}
	\label{tbl:narrator}
	\begin{center}
	\begin{tabular}{l|c c c c c c c} 
		\toprule		
		Method\hspace{12ex} & ~CIDEr~ & ~METEOR~ & ~ROUGE-L~ & ~BLEU-1~ & ~BLEU-2~ & ~BLEU-3~ & ~BLEU-4~ \\
		\midrule
		Uniform& 89.9 &18.3 &28.0&65.7 &40.6 &23.2 &13.2 \\	
		SeqDPP~\cite{Gong_NIPS_2014}& 91.6 &18.3 &28.3& 66.3&41.0&23.6&13.1\\	
		Submodular~\cite{Gygli_CVPR_2015}& 92.0 &18.4 &28.1&66.4&41.0&23.8&13.3\\		
		vsLSTM~\cite{Zhang_ECCV_2016}&92.4&18.2 &28.2 &66.6 &41.5 &24.1 &13.6  \\
		\midrule
		Narrator IoU&93.7 & 18.6 & 28.2 &67.9&42.1 &24.7 &14.1 \\
		Narrator w/o $\Vec{x}$&93.4 & 18.5 & 28.3 &67.3&41.6 &24.5 &14.2 \\
		Narrator w/o $f_l$&96.1 & 19.0 & 29.1 &68.6&42.9 &25.2 &14.5 \\
		Narrator &\bf{98.4} & \bf{19.6} & \bf{29.5} &\bf{69.1}&\bf{43.0} &\bf{25.3} &\bf{15.0} \\
		\bottomrule
	\end{tabular}
	\end{center}
\end{table*}	

%% file: table/tbl_userstudy.tex
\begin{table}[!t]
	\caption
		{
		\small	
			User study results. Numbers indicate the percentage of pairwise preference
		}
	\label{tbl:userstudy}
	\vspace{-2ex}
	\begin{center}
	\begin{tabular}{l|c}
		\toprule
		Compared methods & Percentage of preference\\
		\midrule
		EMB $>$ H-RNN & 70.0\% \\
		ResNarrator $>$ H-RNN & 88.8\% \\
		ResNarrator $>$ EMB & 80.0\% \\
		\midrule
		ResNarrator $>$ GT & 38.1\% \\
		\bottomrule
	\end{tabular}	
	\end{center}
	\vspace{-3ex}
\end{table}	

%% file: sec_conclusion.tex
\section{Conclusion}
\label{sec:conclusion}

In this work,
we have studied the problem of video storytelling.
To address the challenges posed by the diversity of the story and the length and complexity of the video,
we propose a ResBRNN model for context-aware multimodal embedding learning,
and a Narrator model directly optimized on the story for clip selection.
We evaluate our method on a new Video Story dataset,
and demonstrate the efficacy of the proposed method both quantitatively and qualitatively.
One limitation of our method is that the story is limited by sentences in the training set.
For future work, we intend to utilize sentences in-the-wild to further improve the diversity of the story.
We also intend to explore NLP based methods to refine the story with smoother transition between sentences.
Furthermore,
we plan to incorporate recent advances in text representations~\cite{bert} and hard negatives~\cite{VSE} into our framework.


%% file: sec_acknowledgement.tex
This research is supported by the National Research Foundation, Prime Minister's Office, Singapore under its Strategic Capability Research Centres Funding Initiative.

%% file: main.bbl
\begin{thebibliography}{10}
\providecommand{\url}[1]{#1}
\csname url@samestyle\endcsname
\providecommand{\newblock}{\relax}
\providecommand{\bibinfo}[2]{#2}
\providecommand{\BIBentrySTDinterwordspacing}{\spaceskip=0pt\relax}
\providecommand{\BIBentryALTinterwordstretchfactor}{4}
\providecommand{\BIBentryALTinterwordspacing}{\spaceskip=\fontdimen2\font plus
\BIBentryALTinterwordstretchfactor\fontdimen3\font minus
  \fontdimen4\font\relax}
\providecommand{\BIBforeignlanguage}[2]{{%
\expandafter\ifx\csname l@#1\endcsname\relax
\typeout{** WARNING: IEEEtran.bst: No hyphenation pattern has been}%
\typeout{** loaded for the language `#1'. Using the pattern for}%
\typeout{** the default language instead.}%
\else
\language=\csname l@#1\endcsname
\fi
#2}}
\providecommand{\BIBdecl}{\relax}
\BIBdecl

\bibitem{Krause_2017_CVPR}
J.~Krause, J.~Johnson, R.~Krishna, and L.~Fei-Fei, ``A hierarchical approach
  for generating descriptive image paragraphs,'' in \emph{{CVPR}}, 2017, pp.
  317--325.

\bibitem{Liang_ICCV_2017}
X.~Liang, Z.~Hu, H.~Zhang, C.~Gan, and E.~P. Xing, ``Recurrent topic-transition
  {GAN} for visual paragraph generation,'' in \emph{{ICCV}}, 2017, pp.
  3362--3371.

\bibitem{Rohrbach_GCPR_2014}
A.~Rohrbach, M.~Rohrbach, W.~Qiu, A.~Friedrich, M.~Pinkal, and B.~Schiele,
  ``Coherent multi-sentence video description with variable level of detail,''
  in \emph{German Conference on Pattern Recognition}, 2014, pp. 184--195.

\bibitem{Yu_CVPR_2016}
H.~Yu, J.~Wang, Z.~Huang, Y.~Yang, and W.~Xu, ``Video paragraph captioning
  using hierarchical recurrent neural networks,'' in \emph{{CVPR}}, 2016, pp.
  4584--4593.

\bibitem{Ranjay_ICCV_2017}
R.~Krishna, K.~Hata, F.~Ren, L.~Fei-Fei, and J.~C. Niebles, ``Dense-captioning
  events in videos,'' in \emph{{ICCV}}, 2017, pp. 706--715.

\bibitem{Xu_TCSVT_2018}
N.~Xu, A.~Liu, Y.~Wong, Y.~Zhang, W.~N.~Y. Su, and M.~S. Kankanhalli,
  ``Dual-stream recurrent neural network for video captioning,'' \emph{{IEEE}
  Transactions on Circuits and Systems for Video Technology}, 2018.

\bibitem{Liu_Access_2018}
A.~Liu, Y.~Qiu, Y.~Wong, Y.~Su, and M.~S. Kankanhalli, ``A fine-grained
  spatial-temporal attention model for video captioning,'' \emph{{IEEE}
  Access}, vol.~6, pp. 68\,463--68\,471, 2018.

\bibitem{Liu_CVIU_2017}
A.~Liu, N.~Xu, Y.~Wong, J.~Li, Y.~Su, and M.~S. Kankanhalli, ``Hierarchical
  {\&} multimodal video captioning: Discovering and transferring multimodal
  knowledge for vision to language,'' \emph{Computer Vision and Image
  Understanding}, vol. 163, pp. 113--125, 2017.

\bibitem{Park_NIPS_2015}
C.~C. Park and G.~Kim, ``Expressing an image stream with a sequence of natural
  sentences,'' in \emph{Advances in Neural Information Processing Systems},
  2015, pp. 73--81.

\bibitem{Liu_AAAI_2017}
Y.~Liu, J.~Fu, T.~Mei, and C.~W. Chen, ``Let your photos talk: Generating
  narrative paragraph for photo stream via bidirectional attention recurrent
  neural networks,'' in \emph{{AAAI}}, 2017, pp. 1445--1452.

\bibitem{Dufaus_ICIP_2000}
F.~Dufaux, ``Key frame selection to represent a video,'' in \emph{{ICIP}},
  2000, pp. 275--278.

\bibitem{Goldman_ACMTG_2006}
D.~B. Goldman, B.~Curless, D.~Salesin, and S.~M. Seitz, ``Schematic
  storyboarding for video visualization and editing,'' \emph{{ACM} Transactions
  on Graphics}, vol.~25, no.~3, pp. 862--871, 2006.

\bibitem{Gygli_CVPR_2015}
M.~Gygli, H.~Grabner, and L.~J.~V. Gool, ``Video summarization by learning
  submodular mixtures of objectives,'' in \emph{{CVPR}}, 2015, pp. 3090--3098.

\bibitem{Lu_CVPR_2013}
Z.~Lu and K.~Grauman, ``Story-driven summarization for egocentric video,'' in
  \emph{{CVPR}}, 2013, pp. 2714--2721.

\bibitem{Potapov_ECCV_2015}
D.~Potapov, M.~Douze, Z.~Harchaoui, and C.~Schmid, ``Category-specific video
  summarization,'' in \emph{ECCV}, ser. Lecture Notes in Computer Science, vol.
  8694, 2014, pp. 540--555.

\bibitem{Zhang_CVPR_2016}
K.~Zhang, W.~Chao, F.~Sha, and K.~Grauman, ``Summary transfer: Exemplar-based
  subset selection for video summarization,'' in \emph{{CVPR}}, 2016, pp.
  1059--1067.

\bibitem{Song_CVPR_2015}
Y.~Song, J.~Vallmitjana, A.~Stent, and A.~Jaimes, ``Tvsum: Summarizing web
  videos using titles,'' in \emph{{CVPR}}, 2015, pp. 5179--5187.

\bibitem{Huang_NAACL_2016}
F.~Huang, Ting-Hao K.and~Ferraro, N.~Mostafazadeh, I.~Misra, J.~Devlin,
  A.~Agrawal, R.~Girshick, X.~He, P.~Kohli, D.~Batra \emph{et~al.}, ``Visual
  storytelling,'' in \emph{{NAACL}}, 2016.

\bibitem{Dai_ICCV_2017}
B.~Dai, D.~Lin, R.~Urtasun, and S.~Fidler, ``Towards diverse and natural image
  descriptions via a conditional {GAN},'' in \emph{{ICCV}}, 2017, pp.
  2970--2979.

\bibitem{Li_HLT_2016}
J.~Li, M.~Galley, C.~Brockett, J.~Gao, and B.~Dolan, ``A diversity-promoting
  objective function for neural conversation models,'' in \emph{{NAACL} {HLT}},
  2016, pp. 110--119.

\bibitem{Mao_CoRR_2014}
J.~Mao, W.~Xu, Y.~Yang, J.~Wang, and A.~L. Yuille, ``Deep captioning with
  multimodal recurrent neural networks (m-{RNN}),'' in \emph{{ICLR}}, 2015.

\bibitem{Xu_ICML_2015}
K.~Xu, J.~Ba, R.~Kiros, K.~Cho, A.~C. Courville, R.~Salakhutdinov, R.~S. Zemel,
  and Y.~Bengio, ``Show, attend and tell: Neural image caption generation with
  visual attention,'' in \emph{{ICML}}, 2015, pp. 2048--2057.

\bibitem{Karpathy_CVPR_2015}
A.~Karpathy and F.~Li, ``Deep visual-semantic alignments for generating image
  descriptions,'' in \emph{{CVPR}}, 2015, pp. 3128--3137.

\bibitem{Junnan_MM}
J.~Li, Y.~Wong, Q.~Zhao, and M.~S. Kankanhalli, ``Attention transfer from web
  images for video recognition,'' in \emph{ACM Multimedia}, 2017, pp. 1--9.

\bibitem{Yao_ICCV_2015}
L.~Yao, A.~Torabi, K.~Cho, N.~Ballas, C.~J. Pal, H.~Larochelle, and A.~C.
  Courville, ``Describing videos by exploiting temporal structure,'' in
  \emph{{ICCV}}, 2015, pp. 4507--4515.

\bibitem{Junnan_ICCV}
J.~Li, Y.~Wong, Q.~Zhao, and M.~S. Kankanhalli, ``Dual-glance model for
  deciphering social relationships,'' in \emph{{ICCV}}, 2017, pp. 2669--2678.

\bibitem{Venugopalan_HLT_2015}
S.~Venugopalan, H.~Xu, J.~Donahue, M.~Rohrbach, R.~J. Mooney, and K.~Saenko,
  ``Translating videos to natural language using deep recurrent neural
  networks,'' in \emph{{NAACL} {HLT}}, 2015, pp. 1494--1504.

\bibitem{Venugopalan_ICCV_2015}
S.~Venugopalan, M.~Rohrbach, J.~Donahue, R.~J. Mooney, T.~Darrell, and
  K.~Saenko, ``Sequence to sequence - video to text,'' in \emph{{ICCV}}, 2015,
  pp. 4534--4542.

\bibitem{Donahue_PAMI_2017}
J.~Donahue, L.~A. Hendricks, M.~Rohrbach, S.~Venugopalan, S.~Guadarrama,
  K.~Saenko, and T.~Darrell, ``Long-term recurrent convolutional networks for
  visual recognition and description,'' \emph{{IEEE} Transactions on Pattern
  Analysis and Machine Learning}, vol.~39, no.~4, pp. 677--691, 2017.

\bibitem{Junnan_NIPS}
J.~Li, Y.~Wong, Q.~Zhao, and M.~S. Kankanhalli, ``Unsupervised learning of
  view-invariant action representations,'' in \emph{Advances in Neural
  Information Processing Systems}, 2018, pp. 1260--1270.

\bibitem{Gao_TMM_2018}
L.~Gao, Z.~Guo, H.~Zhang, X.~Xu, and H.~T. Shen, ``Video captioning with
  attention-based {LSTM} and semantic consistency,'' \emph{IEEE Transactions on
  Multimedia}, vol.~19, no.~9, pp. 2045--2055, 2017.

\bibitem{Hodosh_JAIR_2013}
M.~Hodosh, P.~Young, and J.~Hockenmaier, ``Framing image description as a
  ranking task: Data, models and evaluation metrics,'' \emph{Journal of
  Artificial Intelligence Research}, vol.~47, pp. 853--899, 2013.

\bibitem{Ordonez_NIPS_2011}
V.~Ordonez, G.~Kulkarni, and T.~L. Berg, ``Im2text: Describing images using 1
  million captioned photographs,'' in \emph{Advances in Neural Information
  Processing Systems}, 2011, pp. 1143--1151.

\bibitem{Socher_TACL_2014}
R.~Socher, A.~Karpathy, Q.~V. Le, C.~D. Manning, and A.~Y. Ng, ``Grounded
  compositional semantics for finding and describing images with sentences,''
  \emph{{TACL}}, vol.~2, pp. 207--218, 2014.

\bibitem{Xu_AAAI_2015}
R.~Xu, C.~Xiong, W.~Chen, and J.~J. Corso, ``Jointly modeling deep video and
  compositional text to bridge vision and language in a unified framework,'' in
  \emph{{AAAI}}, 2015, pp. 2346--2352.

\bibitem{Zhu_ICCV_2015}
Y.~Zhu, R.~Kiros, R.~S. Zemel, R.~Salakhutdinov, R.~Urtasun, A.~Torralba, and
  S.~Fidler, ``Aligning books and movies: Towards story-like visual
  explanations by watching movies and reading books,'' in \emph{{ICCV}}, 2015,
  pp. 19--27.

\bibitem{Dong_TMM_2018}
J.~Dong, X.~Li, and C.~G.~M. Snoek, ``Predicting visual features from text for
  image and video caption retrieval,'' \emph{{IEEE} Transactions on
  Multimedia}, vol.~20, no.~12, pp. 3377--3388, 2018.

\bibitem{Tsai_ICCV_2017}
Y.~H. Tsai, L.~Huang, and R.~Salakhutdinov, ``Learning robust visual-semantic
  embeddings,'' in \emph{{ICCV}}, 2017, pp. 3591--3600.

\bibitem{Faghri_BMVC_2018}
F.~Faghri, D.~J. Fleet, J.~Kiros, and S.~Fidler, ``{VSE++:} improving
  visual-semantic embeddings with hard negatives,'' in \emph{{BMVC}}, 2018,
  p.~12.

\bibitem{Lu_TMM_2014}
S.~Lu, Z.~Wang, T.~Mei, G.~Guan, and D.~D. Feng, ``A bag-of-importance model
  with locality-constrained coding based feature learning for video
  summarization,'' \emph{IEEE Transactions on Multimedia}, vol.~16, no.~6, pp.
  1497--1509, 2014.

\bibitem{Plummer_CVPR_2017}
B.~A. Plummer, M.~Brown, and S.~Lazebnik, ``Enhancing video summarization via
  vision-language embedding,'' in \emph{{CVPR}}, 2017, pp. 1052--1060.

\bibitem{Varini_TMM_2017}
P.~Varini, G.~Serra, and R.~Cucchiara, ``Personalized egocentric video
  summarization of cultural tour on user preferences input,'' \emph{IEEE
  Transactions on Multimedia}, vol.~19, no.~12, pp. 2832--2845, 2017.

\bibitem{Pablos_TMM_2018}
A.~T. de~Pablos, Y.~Nakashima, T.~Sato, N.~Yokoya, M.~Linna, and E.~Rahtu,
  ``Summarization of user-generated sports video by using deep action
  recognition features,'' \emph{{IEEE} Transactions on Multimedia}, vol.~20,
  no.~8, pp. 2000--2011, 2018.

\bibitem{Xu_2019_CVPR}
B.~Xu, Y.~Wong, J.~Li, Q.~Zhao, and M.~S. Kankanhalli, ``Learning to detect
  human-object interactions with knowledge,'' in \emph{{CVPR}}, 2019.

\bibitem{Vasudevan_MM_2017}
A.~B. Vasudevan, M.~Gygli, A.~Volokitin, and L.~V. Gool, ``Query-adaptive video
  summarization via quality-aware relevance estimation,'' in \emph{{ACM}
  Multimedia}, 2017, pp. 582--590.

\bibitem{Yeung_CVPRW_2014}
S.~Yeung, A.~Fathi, and L.~Fei{-}Fei, ``{VideoSET}: Video summary evaluation
  through text,'' in \emph{{CVPR} Workshop}, 2014.

\bibitem{Sah_WACV_2017}
S.~Sah, S.~Kulhare, A.~Gray, S.~Venugopalan, E.~Prud'hommeaux, and R.~W.
  Ptucha, ``Semantic text summarization of long videos,'' in \emph{{WACV}},
  2017, pp. 989--997.

\bibitem{V2TS}
B.~Chen, Y.~Chen, and F.~Chen, ``Video to text summary: Joint video
  summarization and captioning with recurrent neural networks,'' in
  \emph{{BMVC}}, 2017.

\bibitem{Williams_ML_1992}
R.~J. Williams, ``Simple statistical gradient-following algorithms for
  connectionist reinforcement learning,'' \emph{Machine Learning}, vol.~8, pp.
  229--256, 1992.

\bibitem{Ba_CoRR_2014}
J.~Ba, V.~Mnih, and K.~Kavukcuoglu, ``Multiple object recognition with visual
  attention,'' in \emph{{ICLR}}, 2015.

\bibitem{Mnih_2014_NIPS}
V.~Mnih, N.~Heess, A.~Graves, and K.~Kavukcuoglu, ``Recurrent models of visual
  attention,'' in \emph{Advances in Neural Information Processing Systems},
  2014, pp. 2204--2212.

\bibitem{Liu_ICCV_2017}
S.~Liu, Z.~Zhu, N.~Ye, S.~Guadarrama, and K.~Murphy, ``Improved image
  captioning via policy gradient optimization of spider,'' in \emph{{ICCV}},
  2017, pp. 873--881.

\bibitem{Ranzato_ICLR_2016}
M.~Ranzato, S.~Chopra, M.~Auli, and W.~Zaremba, ``Sequence level training with
  recurrent neural networks,'' in \emph{{ICLR}}, 2016.

\bibitem{Rennie_CVPR_2016}
S.~J. Rennie, E.~Marcheret, Y.~Mroueh, J.~Ross, and V.~Goel, ``Self-critical
  sequence training for image captioning,'' in \emph{{CVPR}}, 2016, pp.
  7008--7024.

\bibitem{Yeung_2016_CVPR}
S.~Yeung, O.~Russakovsky, G.~Mori, and L.~Fei{-}Fei, ``End-to-end learning of
  action detection from frame glimpses in videos,'' in \emph{{CVPR}}, 2016, pp.
  2678--2687.

\bibitem{Lan_CVPR_2018}
S.~Lan, R.~Panda, Q.~Zhu, and A.~K. Roy{-}Chowdhury, ``Ffnet: Video
  fast-forwarding via reinforcement learning,'' in \emph{{CVPR}}, 2018, pp.
  6771--6780.

\bibitem{Kiros_NIPSW_2014}
R.~Kiros, R.~Salakhutdinov, and R.~S. Zemel, ``Unifying visual-semantic
  embeddings with multimodal neural language models,'' in \emph{{NIPS Deep
  Learning Workshop}}, 2014.

\bibitem{Mikolov_NIPS_2013}
T.~Mikolov, I.~Sutskever, K.~Chen, G.~S. Corrado, and J.~Dean, ``Distributed
  representations of words and phrases and their compositionality,'' in
  \emph{Advances in Neural Information Processing Systems}, 2013, pp.
  3111--3119.

\bibitem{Cho_EMNLP_2014}
K.~Cho, B.~van Merrienboer, {\c{C}}.~G{\"{u}}l{\c{c}}ehre, D.~Bahdanau,
  F.~Bougares, H.~Schwenk, and Y.~Bengio, ``Learning phrase representations
  using {RNN} encoder-decoder for statistical smachine translation,'' in
  \emph{{EMNLP}}, 2014, pp. 1724--1734.

\bibitem{Chung_CoRR_2014}
J.~Chung, {\c{C}}.~G{\"{u}}l{\c{c}}ehre, K.~Cho, and Y.~Bengio, ``Empirical
  evaluation of gated recurrent neural networks on sequence modeling,'' in
  \emph{{NIPS Deep Learning Workshop}}, 2014.

\bibitem{He_CVPR_2016}
K.~He, X.~Zhang, S.~Ren, and J.~Sun, ``Deep residual learning for image
  recognition,'' in \emph{{CVPR}}, 2016, pp. 770--778.

\bibitem{Kingma_CoRR_2014}
D.~P. Kingma and J.~Ba, ``Adam: {A} method for stochastic optimization,'' in
  \emph{{ICLR}}, 2015.

\bibitem{Xu_CVPR_2016}
J.~Xu, T.~Mei, T.~Yao, and Y.~Rui, ``{MSR-VTT:} {A} large video description
  dataset for bridging video and language,'' in \emph{{CVPR}}, 2016, pp.
  5288--5296.

\bibitem{Papineni_ACL_2002}
K.~Papineni, S.~Roukos, T.~Ward, and W.~Zhu, ``{BLUE}: a method for automatic
  evaluation of machine translation,'' in \emph{{ACL}}, 2002, pp. 311--318.

\bibitem{Banerjee_ACL_2005}
S.~Banerjee and A.~Lavie, ``{METEOR}: An automatic metric for mt evaluation
  with improved correlation with human judgments,'' in \emph{{ACL} workshop on
  intrinsic and extrinsic evaluation measures for machine translation and/or
  summarization}, 2005, pp. 65--72.

\bibitem{Vedantam_CVPR_2015}
R.~Vedantam, C.~L. Zitnick, and D.~Parikh, ``Cider: Consensus-based image
  description evaluation,'' in \emph{{CVPR}}, 2015, pp. 4566--4575.

\bibitem{Socher_CVPR_2010}
R.~Socher and F.~Li, ``Connecting modalities: Semi-supervised segmentation and
  annotation of images using unaligned text corpora,'' in \emph{{CVPR}}, 2010,
  pp. 966--973.

\bibitem{Lin_ACL_2004}
C.-Y. Lin, ``Rouge: A package for automatic evaluation of summaries,'' in
  \emph{ACL workshop}, 2004.

\bibitem{Chen_CoRR_2015}
X.~Chen, H.~Fang, T.~Lin, R.~Vedantam, S.~Gupta, P.~Doll{\'{a}}r, and C.~L.
  Zitnick, ``Microsoft {COCO} captions: Data collection and evaluation
  server,'' \emph{CoRR}, vol. abs/1504.00325, 2015.

\bibitem{Kilickaya_EACL_2017}
M.~Kilickaya, A.~Erdem, N.~Ikizler{-}Cinbis, and E.~Erdem, ``Re-evaluating
  automatic metrics for image captioning,'' in \emph{{EACL}}, 2017, pp.
  199--209.

\bibitem{Gong_NIPS_2014}
B.~Gong, W.~Chao, K.~Grauman, and F.~Sha, ``Diverse sequential subset selection
  for supervised video summarization,'' in \emph{Advances in Neural Information
  Processing Systems}, 2014, pp. 2069--2077.

\bibitem{Zhang_ECCV_2016}
K.~Zhang, W.~Chao, F.~Sha, and K.~Grauman, ``Video summarization with long
  short-term memory,'' in \emph{ECCV}, ser. Lecture Notes in Computer Science,
  vol. 9911, 2016, pp. 766--782.

\bibitem{bert}
J.~Devlin, M.~Chang, K.~Lee, and K.~Toutanova, ``{BERT:} pre-training of deep
  bidirectional transformers for language understanding,'' \emph{arXiv preprint
  arXiv:1810.04805}, 2018.

\bibitem{VSE}
F.~Faghri, D.~J. Fleet, J.~Kiros, and S.~Fidler, ``{VSE++:} improving
  visual-semantic embeddings with hard negatives,'' in \emph{{BMVC}}, 2018,
  p.~12.

\end{thebibliography}
